\documentstyle[preprint,aps]{revtex}

\begin{document}
\begin{titlepage}
\rightline{AS-ITP-97-22, 1997}
%\leftline{August, 1997}
%
\vspace{3cm}
\begin{center}
{\LARGE\bf A Possible Unification Model \\ for All Basic Forces}
\end{center}
\bigskip
\begin{center}
{\large\bf Wu Yue-Liang$^{\dagger\dagger}$\footnote{Supported 
in part by Outstanding Young Scientist Fund of China} and Zhou Guang-Zhao (K.C. Chou)$^{\dagger}$ } \\
$^{\dagger}$Chinese Academy of Sciences \\ 
             Beijing 100086  P.R. China  \\
 \vspace{0.3cm}
        $^{\dagger \dagger}$Institute of Theoretical Physics \\
       Chinese Academy of Sciences, P.O. Box 2735 \\
      Beijing, 100080, P.R. China 
\end{center}
\vspace{4cm}
\end{titlepage}
\draft
\preprint{AS-ITP-97-22, 1997}  
\title{A Possible Unification Model for All Basic Forces}
\author{ Wu Yue-Liang $^{\dagger\dagger} $\footnote{Supported 
in part by Outstanding Young Scientist Fund of China} and Zhou Guang-Zhao (K.C. Chou)$^{\dagger}$} 
\address{ $^{\dagger}$Chinese Academy of Sciences \\
         Beijing 100086  P.R. China \\
 \vspace{0.3cm}
        $^{\dagger\dagger}$Institute of Theoretical Physics \\
       Chinese Academy of Sciences, P.O. Box 2735 \\
      Beijing, 100080, P.R. China \\ }
%\date{August 1997}
\maketitle

\begin{abstract}
 A unification model for strong, electromagnetic, weak and gravitational 
forces is proposed. The tangent space of ordinary coordinate 
4-dimensional spacetime is a submanifold of an 14-dimensional 
{\it internal spacetime } spanned by four frame fields. The 
unification of the standard model with gravity is governed by 
gauge symmetry in the {\it internal spacetime}. 
\end{abstract}
%\pacs{PACS numbers: 12.10.Gq, 04.50.+h, 11.30.Ly}

\narrowtext

 One of the great theoretical endeavours in this century is to unify 
 gravitational force characterized by the general relativity of Einstein
 \cite{GR} with all other elementary particle forces (strong, 
electromagnetic and weak ) described by Yang-Mills gauge theory\cite{YM}. 
One of the difficulties arises from the no-go theorem \cite{CM} which 
was proved based on a local relativistic quantum field theory in 
4-dimensional spacetime. Most of the attempts to unify all basic forces 
involve higher dimensional spacetime, such as Kaluza-Klein Yang-Mills 
theories\cite{KK}, supergravity theories\cite{SGT} and superstring 
theories\cite{SST}, {\it etc}. In the Kaluza-Klein Yang-Mills theories, 
in order to have a standard model gauge group as the isometry group of 
the manifold, the minimal number of total dimensions has to be 
eleven\cite{W11}. Even so, the Kaluza-Klein approach is not rich 
enough to support the fermionic representations of the standard model 
due to the requirement of the Atiyah-Hirzebruch index theorem. The 
maximum supergravity has SO(8) symmetry, its action is usually also 
formulated as an N=1 supergravity theory in eleven-dimensional 
spacetime. Unfortunately, the SO(8) symmetry is too small to include 
the standard model. Consistent superstring theories have also been built 
based on ten-dimensional spacetime. In superstring theories, all the known 
particle interactions can be reproduced, but millions of vacua have been 
found. The outstanding problem is to find which one is the true vacuum of 
the theory.

  In this note, we will consider an alternative scheme. Firstly, we 
observe that quarks and leptons in the standard model \cite{SM} can 
be unified into a single 16-dimensional representation of complex chiral
spinors in SO(10)\cite{SO10}. Each complex chiral spinor belong to a 
single 4-dimensional representation of SO(1,3). In an unified theory, 
it is an attractive idea to treat these 64 real spinor components on 
the same footing, i.e., they have to be a single representation of a 
larger group. It is therefore natural to consider SO(1,13) as our 
unified group and the gauge potential of SO(1,13) as the fundamental 
interaction that unifies the four basic forces (strong, electromagnetic, 
weak and gravitational) of nature. Secondly, to avoid the restriction 
given by no-go theorem and other problems mentioned above, 
we consider the ordinary coordinate spacetime remains to be an 
4-dimensional manifold $S_{4}$ with metric $g_{\mu\nu}(x)$, $\mu$,
$\nu$=0,1,2,3. At each point P: $x^{\mu}$, there is an d-dimensional 
flat space $M_{d}$ with $d > 4$ and signature (1, -1, $\cdots$, -1). 
We assume that the tangent space $T_{4}$ of $S_{4}$ at point P to be 
an 4-dimensional submanifold of $M_{d}$ spanned by four 
vectors $e_{\mu}^{A}(x)$ $\mu$=0,1,2,3; $A=0, 1, \cdots, d-1$  such that 
\begin{equation}
g_{\mu\nu}(x) = e_{\mu}^{A}(x)e_{\nu}^{A}(x)\eta_{AB}
\end{equation}
where $\eta_{AB} = diag.(1, -1, \cdots, -1)$ can be considered as the 
metric of the flat space $M_{d}$. We shall call $e_{\mu}^{A}(x)$ to 
be the generalized vierbein fields or simply the frame fields. Once 
the the frame fields $e_{\mu}^{A}(x)$ are given, we can always 
supplement with another d-4 vector fields $e_{m}^{A}(x)\equiv 
e_{m}^{A}(e_{\mu}^{A}(x))$, $m =1,2, \cdots , d-4$, such that
\begin{equation}
e_{\mu}^{A}(x)e_{m}^{B}(x)\eta_{AB} = 0, \qquad
e_{m}^{A}(x)e_{n}^{B}(x)\eta_{AB} = g_{mn}
\end{equation}
where $g_{mn} = diag. (-1, \cdots, -1)$. $e_{m}^{A}(x)$ can be 
uniquely determined up to an SO(d-4) rotation. In the flat 
manifold $M_{d}$ we can use $e_{\mu}^{A}(x)$ and $e_{m}^{A}(x)$ 
to decompose it into two orthogonal manifolds $T_{4}\otimes C_{d-4}$. 
Where $C_{d-4}$ will consider to be the {\it internal space} 
describing SO(d-4) internal symmetry besides the spin and is 
spanned by the d-4 orthonormal vectors $e^{A}_{m}(x)$. In the 
new frame system of $M_{d}$ the metric tensor is of the form
\begin{equation}
\left( \begin{array}{cc}
g_{\mu\nu}(x) & 0 \\
0 & g_{mn} 
\end{array} \right)
\end{equation}

  With $e_{\mu}^{A}(x)$ and $e_{m}^{A}(x)$, we can now define the 
covariant vectors as $e^{\mu}_{A}(x)$ and $e^{m}_{A}(x)$ satisfying 
\begin{eqnarray}
& & e^{\mu}_{A}(x)e^{A}_{\nu}(x) = g^{\mu}_{\nu}, \qquad e^{m}_{A}(x) 
e^{A}_{n}(x) = g^{m}_{n} \\
& & e^{\mu}_{A}(x)e^{A}_{m}(x) = 0, \qquad e^{m}_{A}(x)e^{A}_{\mu}(x) 
= 0  \nonumber 
\end{eqnarray}
Under general coordinate transformations and the rotations in $M_{d}$, 
$e_{\mu}^{A}(x)$ transform as a covariant vector in ordinary coordinate 
spacetime and a vector in the $M_{d}$ rotation, $e_{m}^{A}(x)$ transform 
as a covariant vector in the $C_{d-4}$ rotation and a vector in 
the $M_{d}$ rotation. For a theory to be invariant under both general 
coordinate transformations and local rotations in the flat 
space $M_{d}$, it is necessary to introduce affine connection 
$\Gamma_{\mu\nu}^{\rho}(x)$ for general coordinate transformations 
and gauge potential $\Omega_{\mu}^{AB}(x)= - \Omega_{\mu}^{BA}(x)$ 
for d-dimensional rotation SO(1,d-1) in $M_{d}$. These transformations 
are connected by the requirement that $T_{4}$ has to be the 
submanifold of $M_{d}$ spanned by four vectors $e_{\mu}^{A}(x)$ 
at point P and $e_{\mu}^{A}(x)$ should be a covariantly constant 
frame and satisfy the condition
\begin{equation} 
D_{\mu}e_{\rho}^{A} = \partial_{\mu}e_{\rho}^{A} 
-\Gamma_{\mu  \rho}^{\sigma}e_{\sigma}^{A} + 
g_{U}\Omega_{\mu B}^{A}e_{\rho}^{B} = 0
\end{equation}
It is then easily verified that   
\begin{eqnarray}
D_{\mu}g_{\rho\sigma} & = & \partial_{\mu} g_{\rho\sigma} 
- \Gamma_{\mu\rho}^{\lambda}g_{\lambda\sigma} 
- \Gamma_{\mu\sigma}^{\lambda}g_{\rho\lambda}= 0  \\
D_{\mu}e^{\rho}_{A} & = & \partial_{\mu}e^{\rho}_{A} + 
\Gamma_{\mu \sigma}^{\rho}e^{\sigma}_{A} - 
g_{U}\Omega_{\mu A}^{B}e^{\rho}_{B} = 0 
\end{eqnarray}

 With the above considerations, we can now construct an invariant 
action under general coordinate transformations in the ordinary 
coordinate spacetime and the local  SO(1,d-1) group symmetry in 
$M_{d}$ with eq. (5) as a constraint. In addition, the action is 
required to have no dimensional parameters and to be renormalizable 
in the sense of the power counting. The general form of the 
action which satisfies these requirements is  
\begin{eqnarray}
S_{B} & = & \int d^{4}x \sqrt{-g} \{   - \frac{1}{4}  F_{\mu\nu}^{AB} 
F_{\rho\sigma}^{CD}g^{\mu\rho}g^{\nu\sigma} \eta_{AC}\eta_{BD} \nonumber \\  
& - & \frac{1}{2}\xi \phi^{2} F_{\mu\nu}^{AB} e^{\mu}_{A}e^{\nu}_{B} 
+ \frac{1}{2}g^{\mu\nu}\partial_{\mu}\phi \partial_{\nu}\phi  
+ \frac{1}{4}\lambda \phi^{4} \\
& + & \zeta F_{\mu\nu}^{AB}F_{\rho\sigma}^{CD}g^{\mu\rho}\eta_{AC} 
e^{\nu}_{B}e^{\sigma}_{D} 
+ a_{1}F_{\mu\nu}^{AB}F_{\rho\sigma}^{CD} e^{\mu}_{C} e^{\nu}_{D} 
e^{\rho}_{A}e^{\sigma}_{B} \nonumber \\
& + & a_{2}F_{\mu\nu}^{AB}F_{\rho\sigma}^{CD}e^{\mu}_{C}e^{\nu}_{B} 
e^{\rho}_{A}e^{\sigma}_{D} + a_{3} F_{\mu\nu}^{AB}F_{\rho\sigma}^{CD}
e^{\mu}_{A}e^{\nu}_{B}e^{\rho}_{C}e^{\sigma}_{D} \} \nonumber 
\end{eqnarray}
where $\phi(x)$ is a scalar field introduced to avoid the dimensional 
coupling constants. $a_{i}$ (i=1,2,3), $\zeta$, $\xi$ and $\lambda$ 
are dimensionless parameters. $F_{\mu\nu}^{AB}$ is the field 
strength defined in a standard way 
\begin{equation}
F_{\mu\nu}^{AB} = \partial_{\mu}\Omega_{\nu}^{AB} 
- \partial_{\nu}\Omega_{\mu}^{AB} + g_{U}(\Omega_{\mu C}^{A} 
\Omega_{\nu}^{CB} - \Omega_{\nu C}^{A} \Omega_{\mu}^{CB})
\end{equation}
The tensor $F_{\mu}^{A}$ is defined as $F_{\mu}^{A} = F_{\mu\nu}^{AB}  
e^{\nu}_{B}$

  Using the frame fields $e^{\mu}_{A}(x)$ and $e^{m}_{A}(e^{A}_{\mu}(x))$, 
we can decompose $\Omega_{\mu}^{AB}(x)$ into three parts $e^{\sigma}_{A}(x) 
\Omega_{\mu}^{AB}(x)e^{\rho}_{B}(x)$ ($\rho$, $\sigma$ = 0,1,2,3 ) 
which describe the gravity, and $e^{m}_{A}(x) \Omega_{\mu}^{AB}(x) 
e^{n}_{B}(x)$ which characterize gauge interactions, as well as 
$e^{m}_{A}(x) \Omega_{\mu}^{AB}(x) e^{\sigma}_{B}(x)$ which connect 
gravity with gauge interactions. From the constraints of eq.(5), we obtain   
\begin{eqnarray}
& & g_{U}e_{\sigma A}(x)\Omega_{\mu}^{AB}(x)e^{\rho}_{B}(x)= 
\Gamma_{\mu\sigma}^{\rho} - e_{\sigma A}\partial_{\mu} e^{\rho A}  \\
& & g_{U}e_{m A}(x)\Omega_{\mu}^{AB}(x)e^{\sigma}_{B}(x)= - e_{m A} 
\partial_{\mu} e^{\sigma A} 
\end{eqnarray}
Similarly, we can reexpress $e^{m}_{A}(x) \Omega_{\mu}^{AB}(x) e^{n}_{B}(x)$ 
as 
\begin{equation}
 e^{n}_{A}(x) \Omega_{\mu}^{AB}(x) e^{m}_{B}(x) = A_{\mu}^{mn}(x) 
- \frac{1}{2}( e^{n}_{A} \partial_{\mu} e^{m A} 
- e^{m}_{A} \partial_{\mu} e^{n A})
\end{equation}
where $A_{\mu}^{mn}(x)= - A_{\mu}^{nm}(x)$ (m, n = 1, $\cdots$, d-4) 
is the gauge potential for (d-4)-dimensional rotation SO(d-4) in $C_{d-4}$.

   Note that not all the gauge fields $\Omega_{\mu}^{AB}(x)$ are simply 
new propagating fields due to the constraints $D_{\mu}e_{\rho}^{A}=0$. 
By counting the constraint equations ($4\times 4 \times$d), 
unknowns $\Omega_{\mu}^{AB}(x)$ (with 4d(d-1)/2 degrees of freedom) 
and $e_{\mu}^{A}(x)$ (with $4\times$d degrees of freedom) as well 
as $\Gamma_{\mu\sigma}^{\rho}$ (with 40 degrees of freedom for the 
symmetric parts $\Gamma_{(\mu\sigma)}^{\rho}= \Gamma_{(\sigma\mu)}^{\rho}$ 
and 24 degrees of freedom for antisymmetric 
parts $\Gamma_{[\mu\sigma]}^{\rho}= - \Gamma_{[\sigma\mu]}^{\rho}$ ), 
one sees that besides the antisymmetric parts $\Gamma_{[\mu\sigma]}^{\rho}$, 
the independent degrees of freedom are (4d + 4(d-4)(d-5)/2). These 
independent degrees of freedom coincide with the degrees of freedom of 
the frame fields $e_{\mu}^{A}(x)$ and the gauge fields $A_{\mu}^{mn}(x)$ 
of the group SO(d-4). In addition, the gauge conditions in the 
coset SO(1,d-1)/SO(d-4) lead to additional constraints (4d-10). 
Thus the independent degrees of freedom are reduced to 
(10 + 4(d-4)(d-5)/2) which exactly match with the degrees of 
freedom of the metric tensor $g_{\mu\nu}(x)$ and the gauge fields 
$A_{\mu}^{mn}(x)$ of the group SO(d-4). For d=14, the resulting 
independent degrees of freedom of the fields are sufficient to 
describe the four basic forces. Where the general relativity of 
the Einstein theory is described by the metric tensor. Photon, 
W-bosons and gluons, that mediate the electromagnetic, weak and 
strong interactions respectively, are different 
manifestations of the gauge potential $A_{\mu}^{mn}(x)$ of 
the symmetry group SO(10)\cite{SO10}. The curvature tensor 
$R_{\mu\nu\sigma}^{\rho}$ and the Ricci tensor $R_{\nu\sigma}= 
R_{\mu\nu\sigma}^{\rho} g_{\rho}^{\mu}$ as well as the scalar 
curvature $R = R_{\nu\sigma}g^{\nu\sigma}$ are simply 
related to the field strength $F_{\mu\nu}^{AB}$ via   
$R_{\mu\nu\sigma}^{\rho} = g_{U}F_{\mu\nu}^{AB}e^{\rho}_{A}e_{\sigma B}$, 
$R_{\nu\sigma}= g_{U}F_{\mu\nu}^{AB}e^{\mu}_{A}e_{\sigma B}$ 
and $R = g_{U}F_{\mu\nu}^{AB}e^{\mu}_{A}e^{\nu}_{B}$. It is not 
difficult to check that $R_{\mu\nu}^{AB}R^{\mu\nu}_{AB}= F_{\mu\nu}^{mn} 
F^{\mu\nu}_{mn} + g_{U}^{-2}R_{\mu\rho\nu\sigma} R^{\mu\rho\nu\sigma}$, 
and $R_{\mu}^{A}R^{\mu}_{A} = g_{U}^{-2}R_{\mu\rho}R^{\mu\rho}$. 
Where $F_{\mu\nu}^{mn}(x)$ is the field strength of the gauge 
potential $A_{\mu}^{mn}(x)$ 
\begin{equation}
F_{\mu\nu}^{mn} = \partial_{\mu}A_{\nu}^{mn} - \partial_{\nu}A_{\mu}^{mn} 
+ g_{U}(A_{\mu q}^{m}A_{\nu}^{qn} - A_{\nu q}^{m}A_{\mu}^{qn})
\end{equation}
With these relations, the action $S_{B}$ can be simply reexpressed as 
\begin{eqnarray}
S_{B} & = & \int d^{4}x \sqrt{-g} \{
-\frac{1}{4} F_{\mu\nu}^{mn} F^{\mu\nu}_{mn} 
+ \frac{1}{2}\partial_{\mu}\phi \partial^{\mu}\phi + \frac{1}{4}\lambda 
\phi^{4} -  \frac{1}{2}\xi g_{U}^{-1}\phi^{2} R  \nonumber   \\
& + & g_{U}^{-2}[\ (a_{1}-\frac{1}{4})  R_{\mu\rho\nu\sigma} 
R^{\mu\rho\nu\sigma} + (a_{2}+ \zeta)R_{\mu\rho}R^{\mu\rho} + 
a_{3}R^{2}\ ] \}
\end{eqnarray}
which has the same form as the action of a multiplicatively 
renormalized unified gauge theory including so-called 
$R^{2}$-gravity and a renormalizable scalar matter field 
as well a nonminimal gravitational-scalar coupling.

   In the real world, there exist three generations of quarks 
and leptons. Each generation of the quarks and leptons has 64 
real degrees of freedom. These degrees of freedom will be 
represented by the 64 components of a single Weyl fermion 
$\Psi_{+}(x)$ belonging to the fundamental spinor representation 
of SO(1,13). The action for fermions is given by 
\begin{equation}
S_{F}= \int d^{4}x \sqrt{-g} \{ \frac{1}{2}\bar{\Psi}_{+} 
e^{\mu}_{A}\Gamma^{A}( i\partial_{\mu} 
+ g_{U}\Omega_{\mu}^{BC}\frac{1}{2}\Sigma_{BC} ) \Psi_{+} + h.c.\}
\end{equation}
where $\Sigma_{AB}$ are the generators of the SO(1, d-1) in the 
spinor representations and given by $\Sigma_{AB} =\frac{i}{4}[\Gamma_{A},  
\Gamma_{B}]$. $\Gamma^{A}$ are the gamma matrices that obey 
$\{\Gamma^{A}, \Gamma^{B}\} = 2 \eta^{AB}$. Note that the resulting total action 
$S = S_{B} + S_{F}$ is simple, but it is nontrival for fermionic 
interactions since the gauge potentials $\Omega_{\mu}^{AB}(x)$ are 
related to the independent degrees of freedom $A_{\mu}^{mn}(x)$ 
and  $e_{\mu}^{A}(x)$ by some nontrival relations given in eqs.(10), 
(11) and (12). In particular, the supplemented frame fields 
$e_{m}^{A}(x)$ will have a highly nonlinear dependence on the frame 
fields $e_{\mu}^{A}(x)$.  

 Now let us consider the conservation laws under the general 
coordinate transformations and local rotation SO(1, d-1).  Under 
the local rotation $\Psi_{+}(x) \rightarrow  
e^{-\frac{1}{2}i\omega^{AB}\Sigma_{AB}} \Psi_{+}(x)$, it is not 
difficult to find the conservation law as
\begin{equation}
D_{\mu} (\sqrt{-g}S^{\mu}_{AB}) - \sqrt{-g}T_{[AB]}\equiv  0
\end{equation}
with 
\begin{eqnarray}
S^{\mu}_{AB} & = & g_{U} \frac{1}{4}\bar{\Psi}_{+} e^{\mu}_{C}
\{\Gamma^{C}, \Sigma_{AB}\} \Psi_{+}  \\
T_{[AB]} & = & -i\frac{1}{2}[\bar{\Psi}_{+} e^{\mu}_{A}\Gamma_{B} 
D_{\mu}\Psi_{+} -(D_{\mu}\bar{\Psi}_{+}) e^{\mu}_{A}\Gamma_{B}\Psi_{+}]
\end{eqnarray} 
The general coordinate transformations lead to the well-known 
energy-momentum conservation law as
\begin{equation}
D^{\nu} (\sqrt{-g}T_{\mu\nu}) \equiv \sqrt{-g} F_{\mu\nu}^{AB}S_{AB}^{\nu} 
\end{equation}
with 
\begin{equation}
T_{\mu\nu} = g_{\mu\nu}{\cal L} -i\frac{1}{2}e_{\nu}^{A}[\ \bar{\Psi}_{+}
\Gamma_{A}D_{\mu}\Psi_{+} - (D_{\mu}\bar{\Psi}_{+}) \Gamma_{A}\Psi_{+}\  ]
\end{equation}
Using the covariantly constant frame fileds $e_{\mu}^{A}(x)$, we can 
project $S^{\mu}_{AB}$ and $T_{[AB]}$ into  
\begin{eqnarray}
S^{\mu}_{\rho\sigma} & = & S^{\mu}_{AB}e_{\rho}^{A}e_{\sigma}^{B}, \\
T_{ [\rho\sigma] } & = & T_{[AB]}e_{\rho}^{A}e_{\sigma}^{B} = 
T_{\rho\sigma} - T_{\sigma\rho}
\end{eqnarray}
The angular momentum conservation law becomes
\begin{equation}
D_{\mu} (\sqrt{-g}S^{\mu}_{\rho\sigma}) - \sqrt{-g}T_{[\rho\sigma]}\equiv  0
\end{equation}
It is then easy to show that these two conservation laws (eqs.(19) 
and (23)) are essentially the same as occurs in special relativity  
by noticing the following relations
\begin{equation}
-T_{[\rho\sigma]} = \partial_{\mu}L^{\mu}_{[\rho\sigma]} \equiv 
\partial_{\mu} \left( x_{\rho} T_{\sigma}^{\ \mu} 
- x_{\sigma}T_{\rho}^{\mu} \right)
\end{equation} 
Here $L^{\mu}_{\rho\sigma}$ is the orbital angular momentum 
tensor and $J^{\mu}_{\rho\sigma} \equiv S^{\mu}_{\rho\sigma} 
+ L^{\mu}_{\rho\sigma}$ represents the total angular momentum tensor. 

   From simple ideas we have provided a unification model for 
strong, electromagnetic, weak and gravitational forces and 
constructed the action without dimensional parameters as the 
basis for quantum theory of all the basic forces of the elementary 
particles. Such a theory is conjectured to be multiplicative 
renormalizable though it may remain an effective theory of a 
more fundamental theory. One can find a formal proof of the 
renormalizability of the $R^{2}$-gravity with a scalar field 
in \cite{R2G}. It was known that in the general relativity 
only the Einstein equations have been tested to be in good 
with known experimental data at the classical level.  Thus 
the general relativity of the Einstein theory may be interpreted 
as a classical theory in the low energy limit, so that the 
Einstein-Hilbert and cosmological terms may be induced as a 
result of the low energy limit\cite{R2G,R2G1}. For instance, 
these terms may result from a spontaneous symmetry breaking.
Finally, we would like to comment on the so called unitary 
problem due to the appearance of the higher derivative terms
within the framework of perturbation theory.  The higher derivative
terms become important as the energy scale goes up to near the 
Planck scale, at that scale gravitational interaction becomes strong
so that the treatment by perturbatively expanding the metric fields 
is no longer suitable. From gauge theory points of view, the local Lorentz
group is not a compact group, not all the components of the gauge fields 
are physical one, additional conditions have to be introduced to 
eliminate those unphysical components. This is similar to the case 
of gauge theories in which gauge conditions have been used to eliminate the
unphysical modes (time and longitudinal modes for massless gauge fields). 
Therefore, to solve the so called unitary problem in gravitational interactions, 
a nonperturbative treatment or an alternative approach has to be developed. 
We shall further issue this problem in our future investigations. 

   We hope that the present model has provided us a new insight 
for unifying all the basic forces within the framework of 
quantum field theory. Though the ideas and the resulting model 
are both simple, there remains more theoretical work and 
experimental efforts needed to test whether they are the true 
choice of nature.

%\end{thebibliography}

\end{document}